%% file: main.tex
\documentclass[pdflatex,sn-basic]{sn-jnl}

\input{mypreamble}

\begin{document}

\title{Automated Testing of Prevalent 3D User Interactions in Virtual Reality Applications}


\author*[1]{\fnm{Ruizhen} \sur{Gu}}\email{rgu10@sheffield.ac.uk}

\author[1]{\fnm{José Miguel} \sur{Rojas}}\email{j.rojas@sheffield.ac.uk}

\author[1]{\fnm{Donghwan} \sur{Shin}}\email{d.shin@sheffield.ac.uk}

\affil[1]{\orgdiv{School of Computer Science}, \orgname{The University of Sheffield}, \orgaddress{\city{Sheffield}, \country{UK}}}


\input{sections/0-abstract}

\maketitle

\input{sections/1-introduction}
\input{sections/2-background}
\input{sections/3-motivation}
\input{sections/4-preliminary}
\input{sections/5-graph}
\input{sections/6-approach}
\input{sections/7-evaluation}
\input{sections/8-discussion}

\input{sections/9-conclusion}

\bibliography{references}

\end{document}

%% file: mypreamble.tex
\usepackage[T1]{fontenc}
\usepackage{graphicx}
\usepackage{hyperref}
\usepackage{color}

\usepackage{amsmath,amssymb,amsfonts}
\usepackage{natbib}
\usepackage[most]{tcolorbox}
\usepackage[inline]{enumitem}
\usepackage{booktabs}
\usepackage[table]{xcolor}
\usepackage{subcaption}
\usepackage{lscape}

\newtcolorbox{rqanswer}{
    enhanced,
    colback=gray!20,
}

\usepackage{xspace}
\newcommand\TOOL{\textsc{XRintTest}\xspace}
\newcommand\DATASET{\textsc{XRBench3D}\xspace}

%% file: sections/0-abstract.tex
\abstract{
Virtual Reality (VR) technologies offer immersive user experiences across various domains, but present unique testing challenges compared to traditional software.
Existing VR testing approaches enable scene navigation and interaction activation, but lack the ability to automatically synthesise realistic 3D user inputs (e.g, grab and trigger actions via hand-held controllers).
Automated testing that generates and executes such input remains an unresolved challenge. Furthermore, existing metrics fail to robustly capture diverse interaction coverage.
This paper addresses these gaps through four key contributions. First, we empirically identify four prevalent interaction types in nine open-source VR projects: \emph{fire}, \emph{manipulate}, \emph{socket}, and \emph{custom}.
Second, we introduce the \emph{Interaction Flow Graph}, a novel abstraction that systematically models 3D user interactions by identifying targets, actions, and conditions.
Third, we construct \DATASET, a benchmark comprising ten VR scenes that encompass 456 distinct user interactions for evaluating VR interaction testing.
Finally, we present \TOOL, an automated testing approach that leverages this graph for dynamic scene exploration and interaction execution.
Evaluation on \DATASET shows that \TOOL achieves great effectiveness, reaching 93\% coverage of \emph{fire}, \emph{manipulate} and \emph{socket} interactions across all scenes, and performing 12x more effectively and 6x more efficiently than random exploration.
Moreover, \TOOL can detect runtime exceptions and non-exception interaction issues, including subtle configuration defects.
In addition, the Interaction Flow Graph can reveal potential interaction design smells that may compromise intended functionality and hinder testing performance for VR applications.
}

\keywords{Extended Reality, Software Testing, Model-based Testing}

%% file: sections/1-introduction.tex
\section{Introduction} \label{sec:intro}

Extended reality (XR) encompasses virtual reality (VR), augmented reality (AR), and mixed reality (MR)~\citep{milgram_augmented_1994}.
This work focuses on VR, which creates simulated experiences using 3D displays and positional tracking to create simulated immersive environments with interactive digital content.
VR applications (VR apps) are software programs that enable users to interact with these environments through head-mounted displays (HMDs) such as the Meta Quest 3\footnote{\url{https://www.meta.com/quest/quest-3/}\label{fn:quest3}} and emerging XR glasses~\citep{gu2025XRGlasses}.
VR technology is being adopted increasingly in various domains, such as entertainment, education~\citep{kavanagh2017SystematicReviewVirtual} and  engineering~\citep{Tadeja2020AeroVR}.
VR development platforms have matured considerably; for example, Unity\footnote{\url{https://unity.com/}} has emerged as one of the mainstream platforms for VR development, supporting major VR operating systems including Meta Horizon\footnote{\url{https://developers.meta.com/horizon/}} and Android XR\footnote{\url{https://www.android.com/xr/}}. 

As VR apps span diverse domains and platforms, their development and testing processes have become substantially more complex, highlighting the need for more comprehensive testing approaches~\citep{andradeUnderstandingVRSoftware2020}. 
VR app testing involves various concerns, including functionality and usability, as well as specific aspects such as object placement accuracy~\citep{yang2024AutomaticOraclePrediction} and cybersickness detection~\citep{li2024LessCybersicknessPlease}. 
Our previous systematic mapping study regarding software testing for XR apps found that, among these test concerns, \emph{functionality} and \emph{user interaction} remain the most critical~\citep{gu2025softwaretestingextendedreality}.
This paper focuses on the challenges of testing VR user interactions to ensure correctness and consistency between user input and system responses in 3D environments. 

Unlike conventional 2D mobile apps, VR app testing requires validating complex 3D interactions and exploring spatial scenarios~\citep{gu2025softwaretestingextendedreality}, exposing critical limitations in existing techniques, most of which fall short of systematically traversing the vast input space to identify all the possible user interactions.
This challenge complicates efforts to achieve adequate coverage of VR scenarios, and is amplified by their virtually limitless variations~\citep{deandradeExploitingDeepReinforcement2023}. 
Most existing works in VR app testing, such as \textsc{VRTest}~\citep{wangVRTestExtensibleFramework2022} and \textsc{VRGuide}~\citep{wangVRGuideEfficientTesting2023}, has largely focused on legacy 2D UI elements instead of the rich 3D interactions that characterise modern VR experiences. 
Building on this line of research, our previous work introduced \textsc{IntenXion}~\citep{gu2025XRTestLibrary}, a test automation library capable of simulating realistic 3D XR inputs through automated controller actions.
However, \textsc{IntenXion} specifically targets \emph{test automation} (i.e., automated execution on predefined tests), and does not address the broader challenge of \emph{automated testing}, which encompasses both automated test generation and execution.

Our previous work introduced \TOOL~\citep{gu2025XRTestingTool}, an automated testing tool that addresses these limitations by systematically modelling user interactions within VR apps and automatically synthesising controller-based interactions. 
However, \TOOL still has several limitations:
\begin{enumerate*}[label=(\arabic*)]
    \item it lacks a comprehensive analysis of prevalent VR user interaction patterns across open-source projects;
    \item it supports automated generation only for fundamental VR actions (\emph{trigger} and \emph{grab});
    \item its evaluation lacks robustness and covers only a limited number of subjects.
\end{enumerate*}

To address the first limitation regarding the absence of empirical data on prevalent VR interaction patterns, this paper first adopts a systematic empirical approach to identify common 3D interaction types in open-source VR projects.
Based on results from ten representative VR scenes, we identify four prevalent interaction categories: \emph{fire}, \emph{manipulate}, \emph{socket}, and \emph{custom}. 
Each category captures distinct interaction mechanics, collectively encompassing both simple and complex composite patterns.

To systematically model these interactions, we propose a novel static model called \emph{Interaction Flow Graph} (IFG). It emphasises the \emph{interaction flows} between interactive objects (e.g., controllers, buttons), representing them as nodes and the directed edges between them as interaction pathways.
The IFG builds upon and extends the \emph{XR User Interaction Graph} (XUI Graph) proposed by \citet{gu2025XRTestingTool}.
While the XUI Graph treats each XR user interaction as a single, atomic action, modern XR experiences often involve interactions that are realised through multiple interdependent actions. This limitation restricts the XUI Graph's ability to represent complex, multi-step interactions common in modern XR environments.
To address this limitation, the IFG extends the original structure by allowing each edge to be annotated with a label representing a sequence of interactions. 
Each interaction in this sequence can consist of multiple actions, enabling finer-grained modelling of complex user behaviours. Moreover, actions in the IFG encompass a rich set of types. For instance, trigger and activation actions may exist in both instant and continuous forms.
By capturing both the structural and behavioural aspects of interactions, the IFG provides a more expressive and adaptable foundation for developing VR testing and analysis approaches.

To enhance the capabilities of existing testing tools that automatically evaluate realistic 3D user interactions in VR applications, we substantially extend \TOOL. The previous version of \TOOL supported only simple interactions between the user and a single interactable object, each limited to a single action (i.e., the \emph{grab} and \emph{tigger} interactions).
In contrast, the new version supports interactions involving multiple objects and models each interaction as a composition of multiple actions, enabling more coverage of user behaviours.
Furthermore, we significantly expand the evaluation of \TOOL compared to the previous version:
\begin{enumerate*}[label=(\arabic*)]
    \item We broaden the types of interactions tested, including the \emph{socket} interactions that support multi-object interactions.
    \item We increase the number of evaluation subjects (i.e., VR scenes) from seven to ten, representing a 43\% increase.
    \item We enlarge the total number of evaluated interactions from 279 to 373, a 34\% increase.
\end{enumerate*}


To evaluate our approach, we construct a novel benchmark dataset \DATASET, which comprises ten VR scenes with 456 3D interactions from nine open-source projects. 
\TOOL demonstrates superior effectiveness and efficiency in covering user interactions, achieving 93\% across all subject scenes, with approximately 12 times more effective and six times more efficient than a random testing approach baseline.
In addition, we propose a novel automated oracle that enables \TOOL to detect non-exceptional functional issues automatically. Furthermore, the IFG helps identify potential interaction smells, providing deeper insights into overall interaction design quality.
We make the replication package of our work publicly accessible, including the implementation of \TOOL and \DATASET benchmark at \url{https://github.com/ruizhengu/XRintTest} and \url{https://github.com/ruizhengu/XRBench3D}, respectively.

In summary, this paper makes the following contributions:
\begin{itemize}
    \item \textbf{VR Interaction Categorisation:} We present the first categorisation of 3D interactions through the analysis of open-source VR projects.
    \item \textbf{Interaction Model for VR Testing:} We formulate a novel systematic abstraction for VR testing by introducing the IFG, a model that formally represents the various interactions in VR environments, serving as the foundation for informing VR testing strategies.
    \item \textbf{Dataset:} We construct \DATASET, a novel benchmark comprising \textbf{ten} VR scenes with \textbf{456} 3D interactions. It facilitates standardised evaluation of VR testing approaches, addressing the absence of dedicated resources. 
    \item \textbf{Technique:} We develop a model-based testing technique that leverages IFGs to explore VR scenes and activate interactions automatically. Our tool, \TOOL, achieves 93\% interaction coverage on the prevalent interactions in \DATASET, outperforming a random baseline 12 times in effectiveness and six times in efficiency. Additionally, \TOOL can also detect exceptions, non-exception interaction issues, and potential interaction smells.
\end{itemize}

This paper is structured as follows.
Section~\ref{sec:background} surveys foundational concepts and related work.
Section~\ref{sec:preliminary} presents our empirical analysis of prevalent VR interaction types.
Section~\ref{sec:motivation} provides the motivating example of this work.
Section~\ref{sec:approach:xui_graph} formalises the interaction flow graph and Section~\ref{sec:approach:xui_explorer} details our proposed technique.
Section~\ref{sec:evaluation} presents our empirical evaluation and results. 
Section~\ref{sec:dicsussions} discusses the implications and findings derived from our study, and 
Section~\ref{sec:conclusion} concludes the paper and outlines future work directions.

%% file: sections/2-background.tex
\section{Background}\label{sec:background}


\subsection{VR Development with Unity} \label{sec:background:unity}

In Unity's development paradigm, a VR app is organised in \emph{scenes}, the containers that represent discrete parts of an app. 
Each scene may contain a set of \emph{GameObjects} (GOs for short), which serve as the fundamental building blocks, encapsulating all entities within a virtual environment.
Each GO can contain a set of \emph{components} with configurable properties that define its specific behaviour. For instance, adding a \emph{Rigidbody} component to a GO enables physics-based motion simulation. When combined with a \emph{Collider} component, the GO can physically interact with other objects in the environment.
Notably, besides the use of built-in components, developers can extend GOs' functionality by attaching custom \emph{scripts} that define specific behaviours and interaction patterns. 


\subsubsection{XR Interaction Toolkit}

The XR Interaction Toolkit\footnote{\url{https://docs.unity3d.com/Packages/com.unity.xr.interaction.toolkit@3.1}} (XRI) is a high-level framework within Unity for developing XR experiences.
XRI encompasses three primary functional domains:
\begin{enumerate*}[label=(\arabic*)]
    \item \textbf{3D Interaction:} Facilitates manipulation of GOs through \emph{interactors} and \emph{interactables};
    \item \textbf{UI Interaction:} Providing UI components for interaction with \emph{controls} (i.e., UI elements such as buttons, sliders, dropdown menus); and 
    \item \textbf{Locomotion:} Enables scene navigation through various methods, such as teleportation, continuous movement, and climbing. 
\end{enumerate*}

While \TOOL necessarily involves essential VR functionalities such as locomotion for scene navigation (Section~\ref{sec:approach:xui_explorer}), this work does not specifically address the testing of those aspects. 
Instead, our primary focus is on testing of \emph{3D} user interactions, and we discuss their underlying mechanisms in detail.

\subsubsection{3D Interaction}

\begin{figure}
    \centering
    \includegraphics[width=0.8\linewidth]{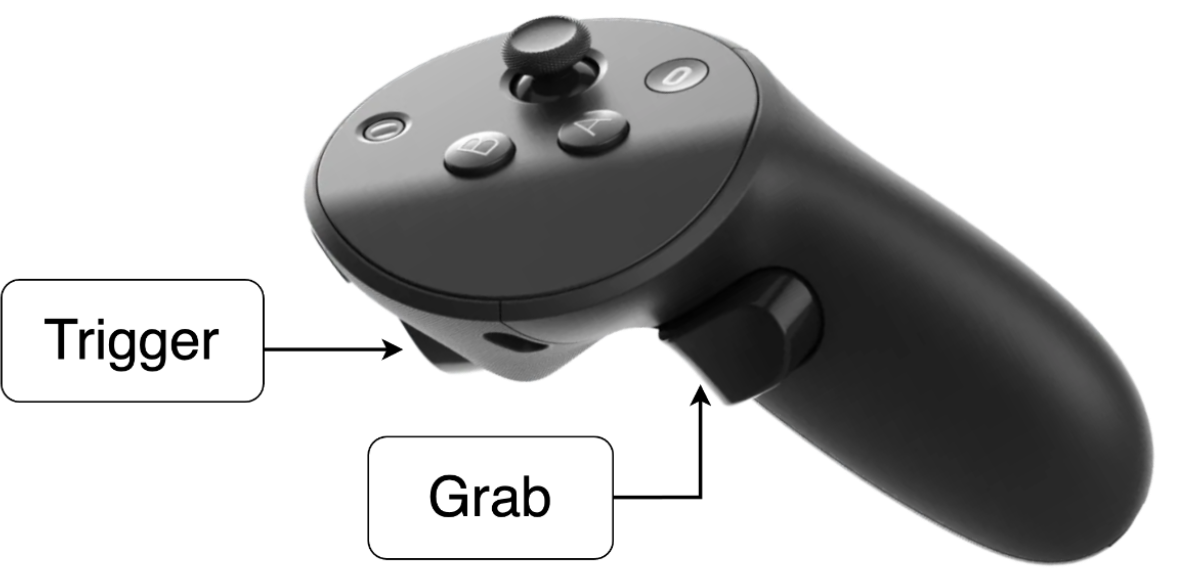}
    \caption{Meta Quest Touch Pro Controller (Right)}
    \label{fig:xr_controller}
\end{figure}

The 3D interaction system forms the core functionality of VR interaction and consists of two main components:
\begin{itemize}
    \item \textbf{Interactors:} \emph{elements} in the scene directly controlled by the user, typically through XR controllers, hand tracking, or touch screens on mobile AR devices. 
    Interactors initiate interactions with interactable objects.
    \item \textbf{Interactables:} \textit{objects} in the scene that the user can interact with. Interactable objects maintain internal states that reflect how they are being interacted with.
\end{itemize}
%
This paradigm is closely related to intuitive engagement with virtual objects through physical hand-held controllers. 
Typical VR controllers, such as the Meta Quest Touch Pro\footnote{\url{https://www.meta.com/quest/accessories/quest-pro-touch-pro-controller/}} displayed in Figure~\ref{fig:xr_controller}, feature two primary buttons designed for 3D interaction: the \emph{trigger} button for actions like firing weapons, and the \emph{grab} button for grasping and manipulating objects.


In the XRI framework, both interactors and interactables are implemented as scripts attached to GOs. The framework provides predefined components to facilitate specific interactions, such as the \texttt{XR Grab Interactable} that implements the basic grab functionality. These components also serve as foundations to implement customised interaction patterns tailored to specific app requirements.
While ``interactor'' and ``interactable'' are specific terms from XRI, we use them consistently throughout this paper to refer to the conceptual input source and target of all VR user interactions, regardless of the development platforms.

\subsubsection{GameObject Layers and Interaction Layers} \label{sec:background:unity:layers}

When using Unity to develop VR applications, two types of \emph{layers} are relevant to interaction: \emph{GameObject Layers} and \emph{Interaction Layers}.

\begin{itemize}
    \item \textbf{GameObject Layers\footnote{\url{https://docs.unity3d.com/6000.3/Documentation/ScriptReference/GameObject-layer.html}}:} Unity’s built‑in layer system, used by subsystems like physics, cameras. A key use is configuring which objects can collide.
    For example, the player avatar and non-player character (NPCs) can be assigned to a ``Player'' layer so they collide with the environment, while UI elements on a ``UI'' layer are excluded from physics collisions.
    \item \textbf{Interaction Layer Masks\footnote{\url{https://docs.unity3d.com/Packages/com.unity.xr.interaction.toolkit@3.1/manual/interaction-layers.html\#interaction-layers-settings}}:} A layer system used by the XRI framework to decide which \emph{Interactors} (e.g., controllers, sockets) can interact with which \emph{Interactables} (e.g., grabbable objects). 
    Interactors and Interactables must share at least one common interaction layer in their Interaction Layer Mask to allow interactions.
    For instance, a controller interactor assigned to the ``Desk'' interaction layer will only interact with interactables that also include the ``Desk'' interaction layer, and will ignore those configured only with a ``Chair'' interaction layer. 
    More details regarding how interaction layer masks affect socket interactions are discussed in Section~\ref{sec:preliminary:discussions}.
\end{itemize}

Misconfigured layers can cause failures in XR testing. GameObject Layers affect whether interactors can physically detect objects, while Interaction Layers determine whether XR interaction can be properly activated (e.g., grabbed).
Automated testing strategies should consider both layer configurations to ensure interaction coverage across different layers.

\subsection{XR Scene Testing} \label{sec:background:scene_testing}

A systematic mapping study for XR app testing conducted by~\citet{gu2025softwaretestingextendedreality} introduced the concept of \emph{scene testing}, which focuses on exploring XR scenes through two core tasks: scene navigation and interaction event triggering.
Wang et al. pioneered VR testing with \textsc{VRTest}~\citep{wangVRTestExtensibleFramework2022} and \textsc{VRGuide}~\citep{wangVRGuideEfficientTesting2023}, the first tools for automated VR scene testing. 
Despite their novel contributions, these approaches rely on and were evaluated using simplified VR projects that \emph{lack comprehensive 3D interaction mechanisms}. 
Specifically, they exclusively adopted basic UI event triggering, instead of the more realistic 3D interaction mechanisms enabled by VR input devices such as hand-held controllers.

While most prior work predominantly focused on testing VR apps through UI interactions, \citet{gu2025XRTestLibrary} proposed \textsc{IntenXion}, a test automation library enabling realistic 3D XR user inputs via simulated controller actions.
It defines \emph{interaction} as a complete exchange between the user and the target interactable to fulfil a specific user intent (e.g., moving an object from position A to B).
An interaction consists of a sequence of discrete \emph{actions}, each representing an individual user step (e.g., pressing a controller button). 
However, it focuses on \emph{test automation} (i.e., solely automated test execution), unable to address \emph{automated testing}.
Motivated by these research gaps, our work advances the domain by introducing a novel automated testing approach for VR apps that systematically generates and executes 3D user interaction via VR input simulation.

\subsection{Model-based Testing} \label{sec:background:model}

Model-based testing (MBT) employs explicit behaviour models to encode the intended behaviours of a system under test, from which test cases are systematically derived and executed \citep{Utting2012model-basedtesting}. This approach provides a structured framework for test generation that aligns with system design specifications.
In GUI testing, MBT is widely adopted as it abstracts complex application behaviours into formal models that serve as a basis for test generation.
However, exhaustive test generation from these models can be costly~\citep{suGuidedStochasticModelbased2017}, leading to guided approaches that focus on specific test targets~\citep{jiangADGEAutomatedDirected, lai2019GoalDrivenExplorationAndroid}.
Notable examples in Android GUI testing include the \emph{window transition graph} \citep{yangStaticWindowTransition2018} and \emph{screen transition graph} \citep{lai2019GoalDrivenExplorationAndroid}.
In video game testing, models such as extended finite state machine (EFSM)~\citep{Ferdous2021ModelPlayGames} and scenario graph~\citep{Ariyure2021kGameHumanlike} represent game states and actions.

Within VR testing, MBT remains a predominant technique \citep{gu2025softwaretestingextendedreality}. For instance, \citet{correasouza2018AutomatedFunctionalTesting} proposed the \emph{requirements flow graph}, which combines \emph{scene graphs} (hierarchical representations of object relationships in virtual environments~\citep{walsh2002understanding}, see Figure~\ref{fig:scene_graph} as an example) with control flow graphs to construct requirement specifications for VR apps.
Building upon these prior works, our approach models 3D user interactions within VR scenes via the interaction flow graph (IFG) to inform test generation. 
Tests generated from the IFG can compose a list of interleaved interactions (e.g., $I_1^{l1}$, $I_1^{l2}$ in Figure~\ref{fig:xui_graph}) across multiple objects within the scene.
This interaction-centric modelling provides a systematic, informative foundation for automated testing strategies that effectively address the testing challenges unique to VR interactions.



%% file: sections/3-motivation.tex
\section{Motivating Example}\label{sec:motivation}

This section motivates the need for a new testing approach and the IFG to model interactions. The discussion begins with why existing XR scene testing research is insufficient for modern VR software, which increasingly relies on realistic, spatial user interactions.

Existing state-of-the-art automated VR scene testing approaches, such as \textsc{VRTest}~\citep{wangVRTestExtensibleFramework2022} and \textsc{VRGuide}~\citep{wangVRGuideEfficientTesting2023}, can systematically explore VR scenes but rely on and were evaluated using relatively simplified VR projects that \emph{lack comprehensive 3D interaction mechanisms}. 
Specifically, they exclusively adopted basic UI-style events (e.g., point-and-click) instead of the more realistic 3D interaction mechanisms enabled by VR input devices such as hand-held controllers.
Consequently, prior work leaves complex 3D user inputs largely unexplored.

From a technical perspective, the key limitation lies in the user interaction paradigm these approaches target. \textsc{VRTest} and \textsc{VRGuide} rely on Unity's \texttt{Event Trigger} mechanism\footnote{\url{https://docs.unity3d.com/Packages/com.unity.ugui@2.0/manual/script-EventTrigger.html}}, designed specifically for handling UI events.
As a result, these approaches only interact with objects functioning as traditional UI elements. 
To elaborate, they target an earlier generation of VR apps where user interactions were limited to performing ``point-and-click'' style interactions. 
The 3D objects in the environments are fundamentally the same as UI buttons, and interaction reduces to dispatching discrete events rather than simulating continuous physical manipulation. 
The evaluation environments used in \textsc{VRTest} and \textsc{VRGuide} reflect this scope, featuring objects with simple on/off states exclusively triggered by the point-and-click interactions.

In contrast, our approach targets modern XR apps built around rich, spatial 3D interactions. These subjects are developed with comprehensive XR interaction frameworks like Unity's XR Interaction Toolkit, where users interact via 6DoF controllers tracked in 3D space. Here, interaction involves grasping, manipulating, and moving virtual objects in a physically plausible manner, rather than merely ``clicking'' them.
The XR apps in the \DATASET benchmark dataset all adopt this modern paradigm. Because \textsc{VRTest} and \textsc{VRGuide} are inherently tied to UI-style event triggering, they achieved 0\% interaction coverage in our preliminary evaluation and are therefore excluded as baselines in this work.

\begin{figure}[ht]
    \vspace{1mm}
    \centering
    \begin{subfigure}[b]{\linewidth}
        \centering
        \includegraphics[width=0.8\linewidth]{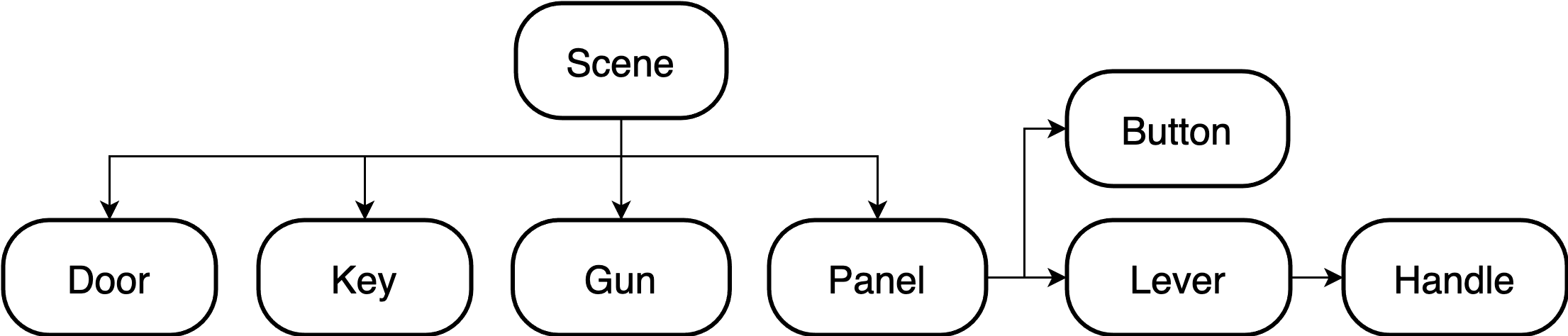}
        \caption{Scene Graph}
        \label{fig:scene_graph}
    \end{subfigure}
    \hfill
    \vspace{11pt}
    \begin{subfigure}[b]{\linewidth}
        \centering
        \includegraphics[width=0.9\linewidth]{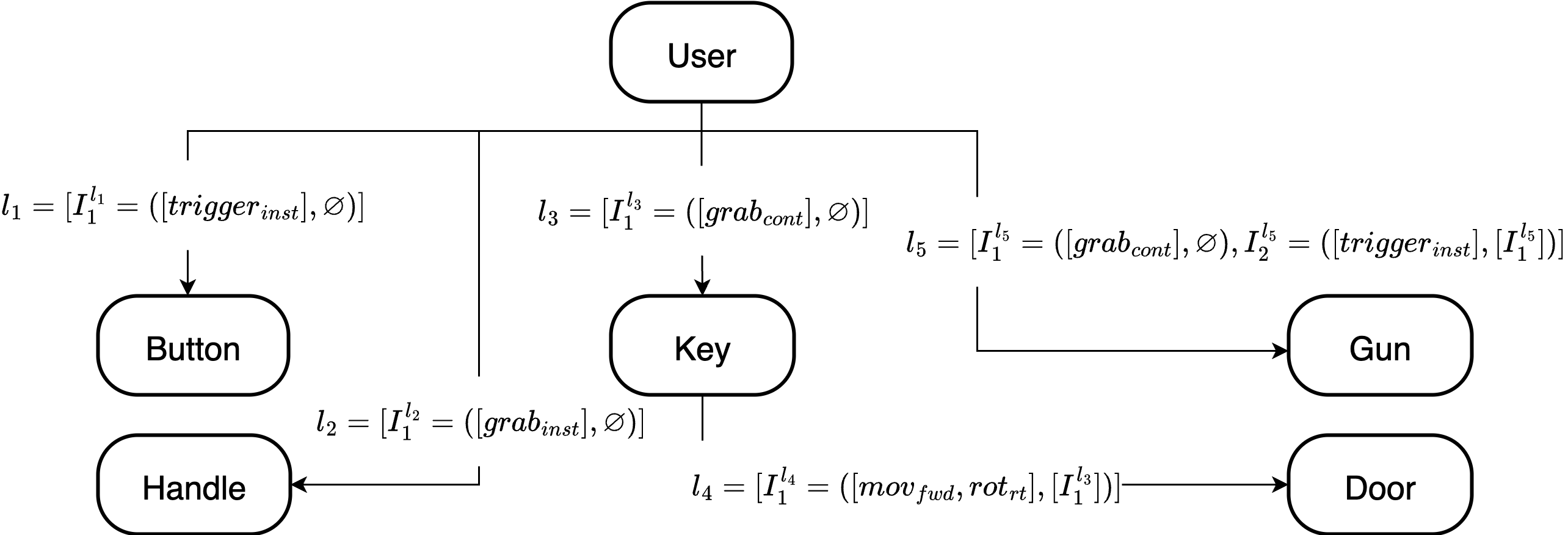}
        \caption{Interaction Flow Graph}
        \label{fig:xui_graph}
    \end{subfigure}
    \caption{Hierarchical Structure and Interaction Flows in VR Scene Modelling \\(Actions: $trigger_{inst}$/$grab_{inst}$ denote instant actions; $trigger_{const}$/$grab_{const}$ denote continuous actions; $mov_{fwd}$ indicates forward controller movement; $rot_{rt}$ indicate rightward controller rotation.)}
    \label{fig:graph_examples}
\end{figure}

Furthermore, the complexity of modern spatial VR interactions necessitates models like the IFG. Figure~\ref{fig:graph_examples} illustrates both the scene graph~\citep{walsh2002understanding} and IFG of an example scene. 
The scene graph (Figure~\ref{fig:scene_graph}) lists all objects in the scene, but not all are interactables, and many support multiple interaction types beyond simple point-and-click.
Different interactables require specific actions to exercise the intended interactions. For instance, the interaction of the ``Handle'' object in the IFG (Figure~\ref{fig:xui_graph}) can only be activated via an instant grab, not other actions like trigger presses.
The IFG captures these interaction dependencies explicitly, enabling automated testing tools to execute the precise actions needed for covering VR interactions.

%% file: sections/4-preliminary.tex
\section{Empirical Analysis of Prevalent VR Interactions} \label{sec:preliminary}

Prior research has highlighted the absence of standardised categories of VR interactions~\citep{gu2025softwaretestingextendedreality, borsting2022SoftwareEngineeringAugmented}. 
While VR input devices typically provide two primary actions, grab and trigger (Section~\ref{sec:background:unity}), the full spectrum of VR interactions remains unclear. 
Existing studies identify common interaction scenarios based on XR design guidelines~\citep{gu2025XRTestLibrary}, but the adoption of various interactions in open-source VR projects is yet to be explored.
To address this gap and support automated VR testing, we adopted an empirical methodology to investigate the prevalent interactions in open-source VR apps.
These insights are critical for prioritising testable interactions and ensuring comprehensive VR functionality coverage.

\subsection{Data Collection}


To collect subjects to analyse prevalent VR interactions, we focus on open-source VR projects built with XRI, a widely adopted framework for implementing 3D VR interactions in Unity-based apps. 
We collected candidate projects from two representative sources, GitHub\footnote{\url{https://github.com/}}, by searching for the XRI package identifier ``\texttt{com.unity.xr.interaction.toolkit}'' and Unity Asset Store\footnote{\url{https://assetstore.unity.com/}}, using the keywords ``XR'' and ``VR''. 
The initial search yielded 46 GitHub repositories and 315 \textbf{free} Unity Asset Store packages (253 VR + 62 XR, with potential duplicates).

To evaluate our approach for testing 3D user interactions in VR apps, our selection criteria focused on identifying \emph{self-contained} projects with sufficient interaction mechanics, such as object manipulation via controllers. We manually screened all candidate projects, first excluding partial components, SDKs, and non-standalone features. We then downloaded, configured, and executed the remaining candidates to verify they contained a minimum of five distinct interactable objects, ensuring a meaningful basis for evaluation. This selection process yielded ten distinct VR scenes across nine qualifying projects (seven from GitHub and two from Unity Asset Store). 


\begin{table}[t]
    \rowcolors{1}{white}{gray!20}
    \centering
    \footnotesize
    \caption{Prevalence of 3D Interactions in VR Projects (Manip.: Manipulate)}
    \begin{tabular}{lllrrrrr} \toprule
    \textbf{ID} & \textbf{Project} & \textbf{Scene} & \textbf{Fire} & \textbf{Manip.} & \textbf{Socket} & \textbf{Custom} &
    \textbf{Tot.} \\ \midrule
         Scene 1 & \href{https://docs.unity3d.com/Packages/com.unity.template.vr@9.1}{VR Template} & Sample Scene & 1 & 8 & 0 & 2 & 11 \\ 
         Scene 2 & \href{https://docs.unity3d.com/Packages/com.unity.xr.interaction.toolkit@3.0/manual/samples-starter-assets.html}{XRI Starter Assets} & Demo Scene & 2 & 6 & 0 & 3 & 11 \\ 
         Scene 3 & \href{https://assetstore.unity.com/packages/tools/game-toolkits/xr-interaction-toolkit-starter-kit-170222}{XRI Starter Kit} & XRI Starter Kit & 10 & 28 & 0 & 30 & 68 \\ 
         Scene 4 & \href{https://github.com/Unity-Technologies/XR-Interaction-Toolkit-Examples}{XRI Examples} & Examples\_Main & 8 & 44 & 5 & 27 & 84 \\ 
         Scene 5 & \href{https://github.com/ValemVR/VR-Game-Jam-Template}{VR-Game-Jam} & Game Scene & 3 & 21 & 2 & 8 & 34 \\ 
         Scene 6 & \href{https://assetstore.unity.com/packages/templates/tutorials/vr-beginner-the-escape-room-163264}{The Escape Room} & Prototype Scene & 6 & 11 & 1 & 2 & 20 \\ 
         Scene 7 & \href{https://assetstore.unity.com/packages/templates/tutorials/vr-beginner-the-escape-room-163264}{The Escape Room} & Escape Room & 19 & 112 & 2 & 7 & 140 \\ 
         Scene 8 & \href{https://github.com/dilmerv/XRToolKitEssentials}{XRToolKitEssentials} & XRSockets & 0 & 5 & 5 & 0 & 10 \\
         Scene 9 & \href{https://github.com/alanplotko/EscapeTheRoomVR}{EscapeTheRoomVR} & SampleScene & 0 & 14 & 0 & 1 & 15 \\
         Scene 10 & \href{https://github.com/MdAsimKhan/VR-Room}{VR-Room} & My VR Room & 6 & 22 & 32 & 3 & 63 \\
         \midrule
        \multicolumn{3}{c}{\textbf{Total}}& 55 & 271 & 47 & 83 & 456 \\
        \bottomrule
    \end{tabular}
\label{tab:xr_projects}
\end{table}

\subsection{Results} \label{sec:preliminary:result}

To identify and categorise these interactions, we manually examined each GO's attributes (e.g., attached interaction scripts) using the Unity Editor. We further validated our findings through hands-on exploration of the scenes using a Meta Quest 3 HMD\footref{fn:quest3}.
Our analysis identified a total of 456 interactions.
During this process, we observed a complexity where interactables were generated or destroyed at runtime. To ensure consistency and reproducibility, we focus only on the interactions that exist in the scene definition. 


Table~\ref{tab:xr_projects} summarises the categorisation of 456 3D interactions across the seven subject scenes into four groups: \emph{fire}, \emph{manipulate}, \emph{socket} and \emph{custom}, each representing various VR interaction mechanisms. 
\emph{Manipulate} interactions are the most straightforward, involving object grabbing via the controller's grab button (Section~\ref{sec:background:unity}), and may also be combined with controller translations for more complex manipulation~\citep{gu2025XRTestLibrary}.
In contrast, \emph{fire} interactions require a composite sequence: the user must first grab and hold a virtual object (e.g., a gun) persistently and then press the trigger button to fire.
This persistent grab distinguishes continuous from instant actions~\citep{gu2025XRTestLibrary}.
Notably, \emph{\textbf{all}} trigger actions in \DATASET occurring with active grabs, underscoring the prevalence of this pattern.

The \emph{socket} and \emph{custom} interactions also adopt composite patterns similar to the \emph{fire} interactions.
\emph{Socket} interactions involve a \texttt{XR Socket Interactor} component, which serves as a dedicated target for a specific interactable, such as the keyhole for a key. 
To complete a socket interaction, the user must continuously grab an object and move it precisely to the socket's position. 
While socket interactions do not directly map to explicit user inputs, they represent intentional responses to user actions, distinguishing them from autonomous behaviours like physics-driven collisions.
The \emph{custom} category includes various interactions tailored to specific scenarios.
For example, a \emph{knob} interaction permits rotational movement along a fixed axis during a continuous grab.
While users typically initiate these interactions through standard grab and trigger actions, successful activation often depends on following additional rules and constraints~\citep{gu2025XRTestLibrary}.
Note that a single object can support multiple interaction types simultaneously. For example, Scene 8 contains 5 interactable objects, each supporting both \emph{manipulate} and \emph{socket} interactions, yielding 10 interactions in total.

Among the 456 interactions, \emph{manipulate} interactions dominate with 271 instances (59\%), followed by \emph{fire} (55, 12\%), \emph{socket} (47, 10\%), and \emph{custom} (83, 18\%) interactions.
These scenes form our benchmark \DATASET. We carefully adapted these projects with minimal changes to ensure compatibility with the latest Unity and XRI versions (Section~\ref{sec:evaluation:setup}). This standardisation enables consistent evaluation across all test subjects while preserving their original interactions.

\subsection{Discussions} \label{sec:preliminary:discussions}

During the empirical analysis of user interactions in the subject VR scenes, we observed several potential issues that can negatively impact the overall testing process and performance.

\subsubsection{Unresponsive Interaction Issues}

During manual exploration of the subject VR scenes, we repeatedly encountered cases where users correctly positioned their controllers and \emph{initiated} the expected interactions, but failed to \emph{activate} them.
Here, activation is defined as successfully issuing the required input on an interactable object so that the associated interaction logic and the object's intended behaviour are fully exercised. 
By contrast, initiation refers to the input sequence that only starts, but does not guarantee that all preconditions for activation are satisfied.
For example, consider a user who positions a controller in contact with a grabbable object and performs the grab action, yet the object fails to respond or be grasped successfully. This case illustrates an interaction that has been initiated but not activated.

We refer to cases where interactions are initiated but fail to activate as \emph{unresponsive interaction issues}. 
In our subject scenes, we identified two concrete instances of such issues, where the target interactables had interaction scripts attached but did not respond to valid interaction inputs.

\subsubsection{Socket Interactions}

\begin{figure}
    \centering
    \includegraphics[width=1\linewidth]{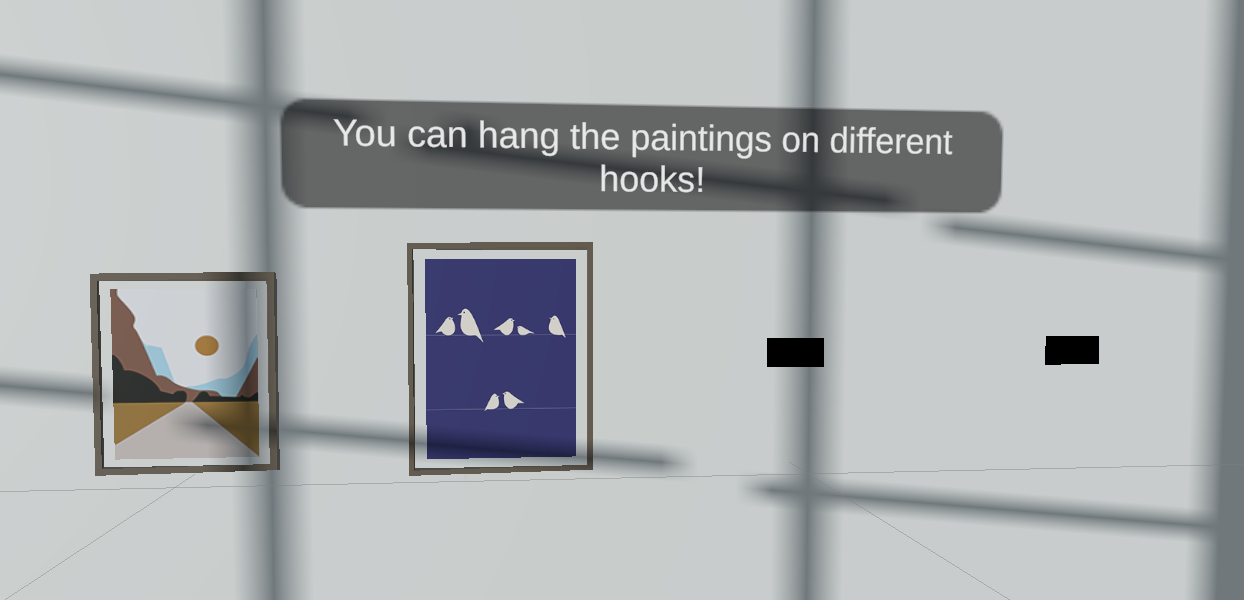}
    \caption{Permutations of Socket Interactions (an example from Scene 10)}
    \label{fig:socket_combination}
\end{figure}

Socket interactions are more complex than \emph{fire} and \emph{manipulate} interactions, as they involve one interactable object and two interactors (e.g., an input interactor and a target socket interactor).
Notably, the Interaction Layer Mask (discussed in Section~\ref{sec:background:unity:layers}) of socket interactors is set to ``Everything'' by default, likely allowing unintended cross-interactions.
For example, while a ``Key'' interactable should only pair with its matching ``Keyhole'' socket interactor, the ``Everything'' layer setting permits any socket interactor to engage with any interactable (e.g., the ``Keyhole'' can be unlocked with a ``Mug''), potentially violating the intended design.
We consider this misconfiguration a potential design smell, which we discuss further in Section~\ref{sec:dicsussions:smell}.

Additionally, one interactable can pair with multiple socket interactors, and vice versa. 
For example, Figure~\ref{fig:socket_combination} shows socket interactions from Scene 10, featuring two ``Painting'' interactables and four ``Picture Hook'' socket interactors. 
When these objects share the same Interaction Layer Mask, fully exercising the socket interaction functionality requires testing all possible pairings between interactables and socket interactors. Since multiple interactables cannot occupy the same socket interactor simultaneously, we calculate the total number of valid socket interactions within one layer mask using the permutation formula $P(n,r) = {}_nP_r = \frac{n!}{(n-r)!}$, where $n$ is the number of interactables and $r$ is the number of socket interactors. In this case, $P(4,2) = \frac{4!}{(4-2)!} = \frac{24}{2} = 12$, yielding 12 distinct socket interactions.

While socket interactors' Interaction Layer Masks default to ``Everything,'' valid socket interactions should restrict each interactor to specific interactables via targeted layers.
To ensure meaningful socket interactions in our subject scenes, we manually screened all socket interactors, inferred their intended target interactables, and assigned corresponding custom layers. For instance, the ``Painting'' interactables and ``Picture Hook'' socket interactors in Scene 10 were grouped within a dedicated ``Painting'' layer, eliminating any ``Everything'' layer defaults.

To compute the total number of socket interactions across a scene, we apply the permutation formula within each Interaction Layer Mask. The total is given by $\sum S = s_1 + s_2 + \dots + s_n$, where $s_i = P(b_i, s_i) = \frac{b_i!}{(b_i - s_i)!}$ represents the number of valid interactions in layer $i$, with $b_i$ interactables and $s_i$ socket interactors.

%% file: sections/5-graph.tex
\section{Interaction Flow Graph} \label{sec:approach:xui_graph}


Modelling 3D user interactions in VR apps is challenging due to their complex, spatial, and highly interactive nature (Section~\ref{sec:background:model}). 
Models such as extended finite state machines (EFSMs)~\citep{Ferdous2021ModelPlayGames} capture states and transitions in detail, but could lead to state‑space explosion~\citep{Amalfitano2015MobiGUITAR} by representing fine‑grained actions (e.g., a discrete controller movement), which hinders their applicability for testing large‑scale VR apps. 

To address these challenges, we propose the Interaction Flow Graph (IFG), a variation of EFSMs that abstracts away low‑level state transitions, represented as a scalable model that can cover all interaction categories in Section~\ref{sec:preliminary}.
In the IFG, nodes represent the user and interactable objects, while directed edges capture valid interactions flowing from source to target. Each edge is annotated with an intended interaction (e.g., a continuous controller movement) as a single transition, regardless of the underlying action sequence. 
This interaction‑centric abstraction provides an intuitive representation of the scene under test (ScUT), enabling systematic test generation and analysis without exhaustively modelling every intermediate state. The IFG thus serves as a repository of testable interactions and their flows, allowing test generation approaches to strategically select and combine interactions to construct meaningful test cases.

We define user interaction flows as \textbf{user-driven exchanges} between interactors and interactables, initiated by intentional user inputs.
This approach explicitly excludes autonomous object interactions (discussed in Section~\ref{sec:preliminary}), which are not user-initiated and thus beyond the scope of this work.
Formally, an IFG is defined as a directed acyclic graph  $G=(N,E,L,I,A,C)$, where: 

\begin{description}
    \item[\textbf{Nodes} ($N$)] represent objects capable of interaction, including both interactors and interactables.
    \textbf{Example:} in the IFG shown in Figure~\ref{fig:xui_graph}, both the interactor \emph{User} and the interactable \emph{Button} are represented as nodes.
    \item[\textbf{Edges} ($E$)] denote valid interactions between nodes.
    An edge $e \in E$ is defined as $e = n_i \rightarrow n_j$, connecting a source $n_i \in N$ to a target $n_j \in N$. 
    The direction of $e$ reflects the flow of interaction from the source to the target node.
    \textbf{Example:} an edge connects the \emph{User} (as the input source) to the \emph{Button} (as the input target).
    \item[\textbf{Labels} ($L$)] annotate edges with corresponding interactions that exercise the functionalities of target interactables.
    Each label $l \in L$ denoting an edge $e$ consists of a set of interactions such that $l = \{I_1, I_2, ..., I_n\}$.
    \textbf{Example:} the label $l_5$ assigned to the edge connecting the \emph{User} and \emph{Gun} consists of two $I_1^{l_5}$ and $I_2^{l_5}$.
    \item[\textbf{Interactions} ($I$)] are the action sequences to form meaningful interactions, possibly conditioned on prerequisites.
    Each interaction $I$ is defined as a tuple $I = (A,C)$, where $A$ represents the action sequence and $C$ is a (possibly empty) set of conditions required to perform the interaction.
    \textbf{Example:} $I_1^{l_2}$ is the only interaction involved in label $l_2$.
    \item[\textbf{Actions} ($A$)] refer to sequences of atomic VR input actions~\citep{gu2025XRTestLibrary}, such as controller movements or grabs.
    The action sequence $A$ is expressed as $A = [a_1, a_2, ..., a_n]$, where each $a \in A$ represents a single atomic action.
    \textbf{Example:} interaction $I_1^{l_4}$ is formed by two atomic actions: $mov_{fwd}$ and $rot_{rt}$ .
    \item[\textbf{Conditions} ($C$)] specify prerequisite \emph{interactions} that must persist to enable the activation of the current interaction. 
    Each condition $c \in C$ references a specific interaction from a particular edge label, which may belong to the same or a different edge.
    For example, an interaction $I_i$ within label $l_q$ could be the sole condition of another interaction $I_j$, expressed as $(I_j=A_j, C_j=[I_i^{l_q}])$.
    A practical illustration is that firing a \emph{Gun} requires the user to have first continuously grabbed the \emph{Gun}.
    \textbf{Example:} interaction $I_2^{l_5}$ within label $l_5$ is guarded by the prior interaction $I_1^{l_5}$.
\end{description}

%
%

Figure~\ref{fig:graph_examples} illustrates both the scene graph~\citep{walsh2002understanding} and IFG of an example scene, clarifying their similarities and distinctions. 
Although our approach does not directly employ scene graphs, we include the scene graph representation in Figure~\ref{fig:scene_graph} to better demonstrate the IFG's distinctive characteristics.
The scene graph (Figure~\ref{fig:scene_graph}) organises objects hierarchically under the entry node \emph{Scene}, with seven children spanning three hierarchical layers.
In contrast, while the IFG (Figure~\ref{fig:xui_graph}) also represents objects within the same scene, it emphasises \emph{interaction flows} rather than hierarchical structure. 
The IFG replaces \emph{Scene} with \emph{User} as the entry node and distils the scene into five nodes connected by five edges with six labels.
These differences highlight three defining properties of an IFG: \emph{interaction prioritisation}, \emph{interaction dependency}, and \emph{interaction brokerage}.


\subsection{Interaction Prioritisation}
In an IFG, each node represents a distinct object capable of interaction, and each edge represents possible interaction flows. This design decouples interaction behaviours from the overall scene hierarchy. 
To enhance precision and reduce complexity, we exclude objects without interaction capabilities, focusing exclusively on those with interaction definitions (i.e., objects with attached scripts that define interactions). 
This approach avoids hierarchical complexity and potential inconsistencies (e.g., misaligned positions between parent and child objects), while enabling more accurate information collection to guide automated testing.

For example, in Figure~\ref{fig:scene_graph}, the \emph{Panel} itself is not interactable, but has two children: an interactable \emph{Button}, and a \emph{Lever}, which is not interactable but contains an interactable child, \emph{Handle}. 
In the IFG shown in Figure~\ref{fig:xui_graph}, only the interactables \emph{Button} and \emph{Handle} are represented as nodes. Non-interactable objects such as \emph{Panel} and \emph{Lever}, although part of the hierarchy, are deliberately omitted.
This selective representation ensures the IFG captures essential interaction information while filtering out non-interactive objects (e.g., for scene management purposes), thereby prioritising meaningful user interactions for testing.

\subsection{Interaction Dependency} \label{sec:approach:xui_graph:dependency}

Interactions may depend on other prerequisite interactions, specified as conditions in edge labels. 
For example, in the IFG in Figure~\ref{fig:xui_graph}, the \emph{Gun} is connected to the \emph{User} with the edge labelled by two interactions.
The first interaction, $I_1^{l_5} = ([grab_{cont}], \varnothing)$, represents continuously grabbing the gun with no prerequisites.
The second interaction, $I_2^{l_5} = ([trigger_{inst}], [I_1^{l_5}])$, corresponds to firing the gun, which requires the persistence of the first interaction.
This dependency ensures $I_2^{l_5}$ could only be activated if the state of $I_1^{l_5}$ persists. 
The IFG explicitly captures such constraints, ensuring correct interaction ordering.

\subsection{Interaction Brokerage} \label{sec:approach:xui_graph:brokerage}

While the \textit{User} typically serves as the default interactor and the starting node of edges in the IFG, our analysis revealed cases where inter-object interactions are initiated by the user, i.e., the socket interactions. 
In such scenarios, an object manipulated by the user acts as a \emph{broker}, mediating interaction with another target object and forming a chained interaction flow (Section~\ref{sec:preliminary}).
For instance, the \emph{Door} and \emph{Key} in Figure~\ref{fig:xui_graph} illustrate this: although the \emph{Door} technically contains a socket interactor~\citep{gu2025XRTestLibrary}, the interaction flow is more accurately modelled with an edge from \emph{Key} to \emph{Door}. 
This represents how the user manipulates the key to unlock the door (rather than the other way around), making the key a broker between the user and the door.
Moreover, interaction $I_1^{l_4}$ (use the key to unlock the door) required the prerequisite interaction $I_1^{l_3}$ (continuous grabbing the key), aligning with our defined interaction dependency property (Section~\ref{sec:approach:xui_graph:dependency}). 

This brokerage mechanism ensures the IFG accurately captures both direct user interactions and indirect, object-mediated interactions, thereby accommodating the multi-mode interaction patterns of complex VR apps.

%% file: sections/6-approach.tex
\section{Automated VR User Interaction Explorer} \label{sec:approach:xui_explorer}

\begin{figure}
    \centering
    \includegraphics[width=\linewidth]{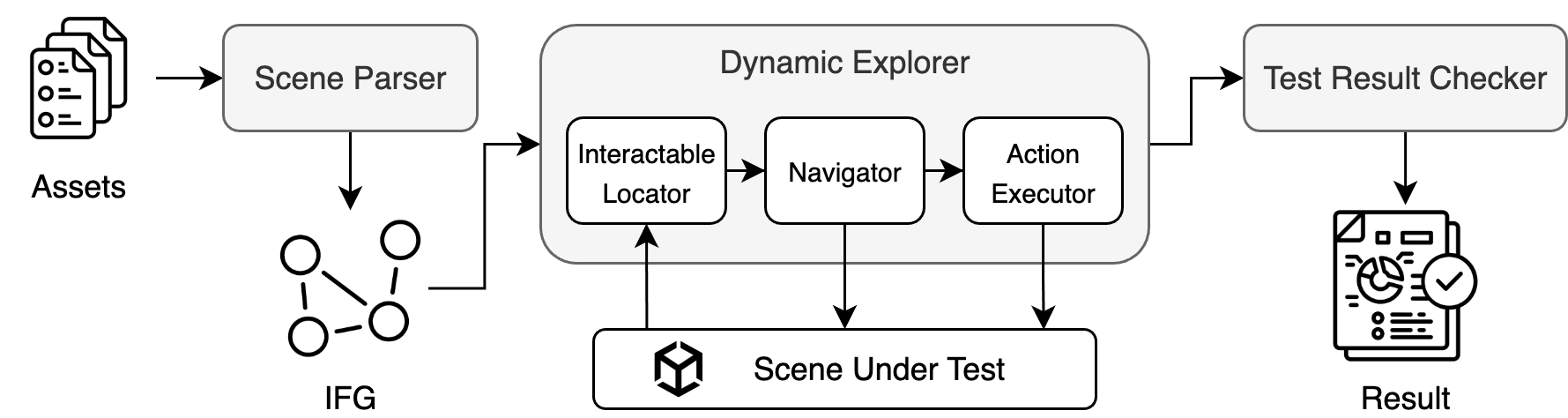}
    \caption{VR User Interaction Testing with \TOOL}
    \label{fig:InteractoBot_apprach}
\end{figure}

In this work, we extend \TOOL, the automated VR user interaction testing tool originally introduced by \citet{gu2025XRTestingTool}. The extended version statically constructs an \emph{Interaction Flow Graph} (IFG) from the VR scene definition, which guides the systematic exploration of the target scene and supports the activation of complex interactions.
Figure~\ref{fig:InteractoBot_apprach} illustrates the overall workflow of \TOOL, which is implemented as a set of scripts within the Unity VR project containing the target scene.
\TOOL takes a Unity project containing the target scene as input.
It first processes the assets required to construct the scene, performing a detailed static analysis to build an IFG representing all interactions within the scene, following the definitions described in Section~\ref{sec:approach:xui_graph}. 
The graph then serves as a guide for the dynamic exploration process, directing the system to engage with the scene's interactables using identified interactions.
%

\subsection{Scene Parser} \label{sec:approach:static}

Unity VR projects comprise source code defining logic and behaviours along with scene definitions that specify GOs and their components.
The objective of the scene parser is to construct a comprehensive IFG (c.f. Section~\ref{sec:approach:xui_graph}) for the ScUT to facilitate the subsequent dynamic exploration phase. 
To ensure accurate interaction identification, we address ambiguities similar to the ``ambiguous GUI'' problem in Android GUI testing~\citep{fulcini2023AnalysisWidgetLayout, arcuschin2022FeasibilityChallengesSynthesizing} where multiple objects in a scene share the identical name. 
We utilise unique identifiers in the IFG, thereby eliminating ambiguity in interactables during dynamic exploration.

We acknowledge that our current approach only analyses static scene definitions and does not detect interactables generated or destroyed at runtime (see Section~\ref{sec:dicsussions:xr}). 
A more comprehensive solution would require extensive analysis of source code and object configurations to identify these dynamic behaviours (Section~\ref{sec:preliminary}). We consider this beyond the present scope, but it is reserved for future research. 

\subsection{Dynamic Explorer} \label{sec:approach:dynamic}

The dynamic explorer's objective is to utilise information from the IFG (Section~\ref{sec:approach:static}) to achieve the two key tasks of VR scene testing: scene navigation and interaction activation (Section~\ref{sec:background:scene_testing}).
This component consists of three processes that communicate with the ScUT in real time: \emph{interactable locator}, \emph{navigator}, and \emph{action executor}. 

\subsubsection{Interactable Locator}

This component leverages the information from the serialised IFG (JSON file; Section~\ref{sec:approach:static}) to precisely identify target GOs for interaction. We adopted a straightforward localisation mechanism that directly maps each interactable name identified in the IFG to its corresponding GO instance at runtime, with the measures to disambiguate interactables in the scene (Section~\ref{sec:approach:static}).

\subsubsection{Navigator} \label{sec:approach:dynamic:navigate}

This navigator handles a core VR scene testing task: scene navigation. It explores the ScUT and positions the simulated user close enough to interactables to activate interactions.
Unlike methods focusing on exploration efficiency~\citep{wangVRGuideEfficientTesting2023}, \TOOL prioritises interaction testing effectiveness.
We adopted a greedy-based navigation algorithm similar to the VRGreedy strategy provided by VRTest~\citep{wangVRTestExtensibleFramework2022}, directing the simulated user towards the closest available interactable (i.e., the closest that has not yet been interacted with) at each step.

Beyond user navigation, the navigator precisely controls the position of simulated input devices to engage interactables using the XR Interaction Simulator~\citep{gu2025XRTestLibrary}.
The simulator cannot directly set camera or controller positions, which only supports indirect movement via simulated inputs (e.g., pressing the ``W'' key to move forward)~\citep{gu2025XRTestLibrary}.
Our implementation automatically generates the corresponding inputs based on the controller’s relative position to the target, producing smooth, natural navigation while maintaining precise interaction control.

\subsubsection{Interaction Executor} \label{sec:approach:dynamic:event}

The interaction executor performs the essential task of activating interactions through synthesised VR inputs. Once the navigator has positioned the controller to contact the target interactable (Section~\ref{sec:approach:dynamic:navigate}), the executor starts engaging with the interactable using the corresponding interactions (i.e., action sequences) identified in the IFG (Section~\ref{sec:approach:static}).

Similar to controller positioning, we simulate specific key events to reproduce controller actions, ensuring consistent executions of various interaction functionalities.
Moreover, our implementation supports both instantaneous and continuous actions, effectively maintaining persistent action states as described in Section~\ref{sec:preliminary}. 

\subsection{Test Result Checker} \label{sec:approach:oracle}


To verify successful interactions during the test sessions, we register listeners to each interactable using XRI's built-in interactable event system~\citep{gu2025XRTestLibrary}, confirming activated interactions. 
Additionally, throughout the testing process, our approach continuously monitors runtime exceptions.
While detecting exceptions is valuable, comprehensive testing approaches should also identify non-exception bugs that compromise user experience without evoking errors~\citep{Su2021functionalfuzzing}.

To address such non-exception bugs, we developed a test oracle to detect subtle problems related to user interactions.
Drawing from the non-responsive issues identified in our empirical analysis (Section~\ref{sec:preliminary:discussions}), this oracle focuses on detecting unresponsive interactions by verifying that correctly initiated interaction sequences successfully activate expected object behaviours.
Our oracle leverages collision detection principles~\citep{Jin2021collisiondetection, deandradeExploitingDeepReinforcement2023}, and is grounded with XRI's requirement that each interactable must possess at least one collider component (Section~\ref{sec:background:unity}). 
We consider that interactions can be activated when the interactable and the controller's colliders intersect.
Our oracle detects valid intersections between them and flags interactions that fail to activate despite intersections, effectively identifying unresponsive interaction issues.

\begin{figure}
    \centering
    \includegraphics[width=0.75\linewidth]{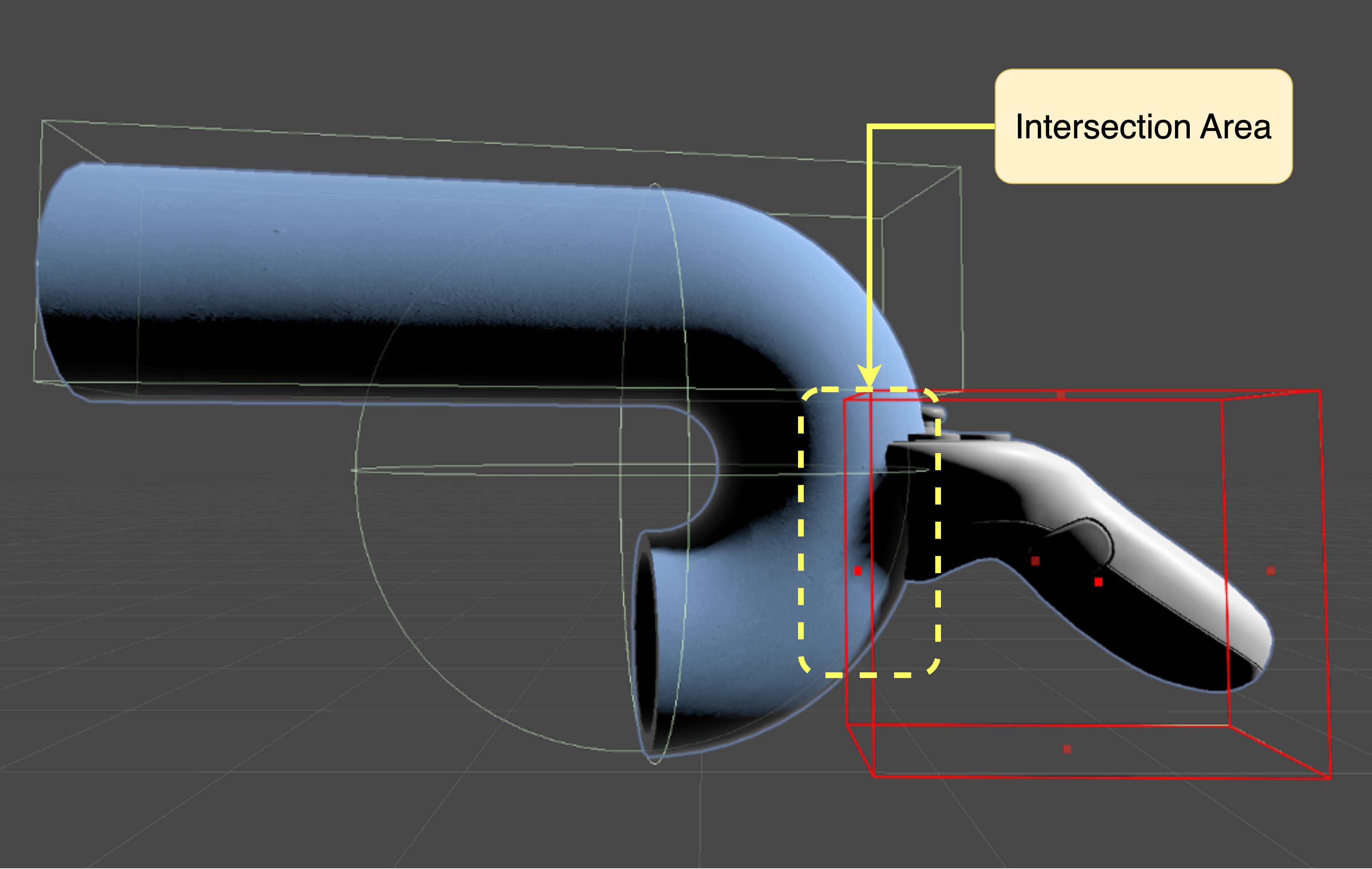}
    \caption{Intersection of Colliders of the Target Interactable (Green Bounds) and Controller (Red Bounds)}
    \label{fig:collider}
\end{figure}

Figure~\ref{fig:collider} illustrates this collision principle. Both the target interactable (the blue stick-shaped object) and the controller incorporate their own colliders. While collider shapes may vary (e.g., box, capsule, sphere), interaction fundamentally depends on the intersection of these colliders. Two objects are considered in contact when their colliders intersect.
Based on this insight, we implement a straightforward oracle for detecting unresponsive interactions: if a valid collision occurs between the controller and the target interactable, but the executed interaction fails to activate the expected behaviours, we deem this as an unresponsive interaction issue in the target interactable. 
This approach effectively identifies objects that appear interactive but do not respond to correct user input, ensuring these subtle issues are systematically detected.

To capture detailed testing results, upon the completion of a test session that attempts to interact with all interactables in the scene, \TOOL generates a test report containing:
\begin{enumerate*}[label=(\arabic*)]
    \item the interaction flow coverage of the scene (detailed in Section~\ref{sec:approach:xui_coverage})
    \item interactables flagged as containing bugs by the collision-based oracle mechanism;
    \item any exceptions encountered during testing.
\end{enumerate*}

\subsection{Interaction Flow Coverage} \label{sec:approach:xui_coverage}

Traditional code coverage metrics offer quantitative insights but are insufficient for assessing Unity-based VR apps, where runtime behaviours depend on both source code and scene configurations~\citep{gu2025XRTestLibrary}. 
Unity's native code coverage primarily measures engine and library code execution (e.g., physics simulations), but cannot ensure activation of expected interactions, even if related scripts are covered.
Video game testing metrics like interactable object coverage~\citep{Coppola2024GameCoverage} quantify the number of activated interactable objects over the total available. 
However, since our IFG emphasises \emph{interactions} rather than objects, these existing metrics are not sufficient for evaluating diverse VR behaviours, as one object can support multiple interactions (Section~\ref{sec:preliminary}).
Covering an object does not imply that all its interactions are covered. For example, if an object supports both \emph{fire} and \emph{manipulate} interactions, executing only one means the object is covered, but its interaction coverage remains incomplete.
Conversely, if all interactions across a scene are covered, full object coverage is inherently achieved.

To systematically evaluate \TOOL's effectiveness, we introduce \emph{Interaction Flow Coverage} ($IFC$), measuring the proportion of activated interactions within a scene.
Unlike object-based metrics, $IFC$ directly captures interaction flow coverage, as a single object can participate in multiple interactions.
We define $IFC$ as the number of interaction flows activated by tests divided by the total flows in the ScUT's IFG.
Our evaluation emphasises the $IFC$ of the prevalent interactions of \emph{fire}, \emph{manipulate}, and \emph{socket} (Section~\ref{sec:preliminary}).

%% file: sections/7-evaluation.tex
\section{Evaluation} \label{sec:evaluation}

To assess the performance of \TOOL, we conducted an evaluation based on the \DATASET benchmark, guided by the following research questions:
\begin{enumerate}[label=\textbf{RQ\arabic*},leftmargin=28pt]
\item 
\textbf{(Coverage):} 
How effectively does \TOOL achieve Interaction Flow Coverage on prevalent VR interactions?
\item 
\textbf{(Efficiency):} 
How efficiently does \TOOL achieve Interaction Flow Coverage on prevalent VR interactions?
\item 
\textbf{(Bug Detection):} 
How well can \TOOL detect bugs in VR apps?
\end{enumerate}

\subsection{Evaluation Setup} \label{sec:evaluation:setup}

The evaluation focuses on \emph{fire} and \emph{manipulate} interactions (Section~\ref{sec:approach:xui_explorer}).
This scope enables valid comparisons, as these standardised interactions are adopted consistently across VR apps. These serve as meaningful targets for automated testing, allowing for fair evaluation of testing performance.
The replication package of our evaluation is publicly accessible, including the implementation of \TOOL and \DATASET benchmark at \url{https://github.com/ruizhengu/XRintTest} and \url{https://github.com/ruizhengu/XRBench3D}, respectively.

\subsubsection{Baseline}

We implemented a baseline variant of \TOOL using a purely random testing strategy, conceptually similar to monkey testing in mobile apps~\citep{Wetzlmaier2016MonkeyGUI}. 
This baseline randomly selects and executes atomic VR actions at 0.1-second intervals, ensuring sufficient action executions for interaction coverage while allowing previous actions to complete.
For each test execution, the simulated user is positioned at a random spawn point to start navigating the scene. To enable composite interactions, we adopted a random key press duration  (0.1 to 0.5 seconds).
Additionally, to prevent testing inefficiencies caused by drifting beyond relevant testing areas, we implemented a position reset function with a weighted probability distribution that favours continuing interactions over position resets. 

We deliberately excluded state-of-the-art VR testing tools, VRTest~\citep{wangVRTestExtensibleFramework2022} and VRGuide~\citep{wangVRGuideEfficientTesting2023}, for methodological validity (detailed in Section~\ref{sec:motivation}). 
Regarding the use of $IFC$ instead of interactable object coverage, interacting with an object does not guarantee all its interactions are covered (Section~\ref{sec:approach:xui_coverage}). Specifically, our subject scenes contain 55 \emph{fire} interactions (Section~\ref{sec:preliminary}), and all of them are associated with \emph{manipulate}. If we only measure interactable objects (e.g., by activating only the \emph{manipulate} interactions), we would miss all 55 \emph{fire} interactions, resulting in incomplete testing coverage despite full object coverage. 

\subsubsection{Running Environment}

To account for the variability of Unity's nondeterministic processes (particularly in physical simulations common to VR), we report the average results from five independent executions of \TOOL and the random baseline for each subject scene. 
This ensures more reliable measurements, despite identical input behaviours that may yield different outcomes. 
Notably, all research questions were addressed using this unified experimental design.

We run our experiments on a MacBook Pro with M3 Pro Chip, 36 GB of RAM, macOS 15.4.1, Unity 6000.0.61f1, and XRI v3.1.1.
To ensure experimental validity, all comparisons used identical settings: a x2 simulation speed and a fixed execution budget of ten simulated minutes (equivalent to five real-time minutes).

\subsection{RQ1: Coverage} \label{sec:evaluation:effectiveness}



\begin{table}[t]
    \scriptsize
    \rowcolors{1}{white}{gray!20}
    \centering
    \caption{Effectiveness of \TOOL on Interaction Flow Coverage}
    \begin{tabular}{c|cccc|cccc} \toprule
    & \multicolumn{4}{c|}{\textbf{Random Baseline}} & \multicolumn{4}{c}{\textbf{\TOOL}} \\
    \textbf{ID} & \textbf{Fire} & \textbf{Manip.} & \textbf{Socket} & \textbf{Total} & \textbf{Fire} & \textbf{Manip.} &
    \textbf{Socket} & \textbf{Total} \\ \midrule
    1 & 0\% & 45\% & N/A & 40\% & 100\% & 100\% & N/A & 100\% \\ 
    2 & 0\% & 43\% & N/A & 33\% & 100\% & 100\% & N/A & 100\% \\ 
    3 & 0\% & 2\% & N/A & 2\% & 100\% & 96\% & N/A & 97\% \\ 
    4 & 0\% & 2\% & 0\% & 2\% & 100\% & 100\% & 100\% & 100\% \\ 
    5 & 0\% & 31\% & 0\% & 28\% & 100\% & 100\% & 100\% & 100\% \\ 
    6 & 0\% & 45\% & 0\% & 29\% & 100\% & 100\% & 100\% & 100\% \\ 
    7 & 0\% & 9\% & 0\% & 8\% & 93\% & 95\% & 100\% & 95\% \\ 
    8 & N/A & 40\% & 0\% & 20\% & N/A & 100\% & 100\% & 100\% \\
    9 & N/A & 0\% & N/A & 0\% & N/A & 100\% & N/A & 100\% \\
    10 & 0\% & 0\% & 0\% & 0\% & 100\% & 91\% & 75\% & 83\% \\
    \midrule
    \textbf{Total} & 0\% & 12\% & 0\% & 8\% & \textbf{97\%} & \textbf{94\%} & \textbf{83\%} & \textbf{93\%} \\ \bottomrule
    \end{tabular}    \label{tab:results:effectiveness}
\end{table}

Table~\ref{tab:results:effectiveness} demonstrates the effectiveness of the random baseline and \TOOL in covering VR user interactions.
The random baseline failed to achieve any successful interaction in the \emph{fire} (0/55, 0\%) and \emph{socket} (0/47, 0\%) interactions, proving particularly challenging for the random approach to address composite interaction patterns.
For \emph{manipulate} interactions, the random baseline achieved only 12\% coverage (31.6/271), resulting in an overall coverage of 8\% (31.6/373).

\TOOL significantly outperformed the random baseline across all subject scenes and interaction categories, achieving 97\% (53.6/55) for \emph{fire}, 94\% (254.6/271) for \emph{manipulate} , and 83\% (39/47) for \emph{socket} interactions, yielding an overall coverage of 93\% (347.2/373). 
The substantial gap highlights the effectiveness of \TOOL compared to random exploration, indicating approximately 12 times greater overall effectiveness. 
The reasons for missed interactions and the relatively lower \emph{socket} interaction coverage will both be discussed in Section~\ref{sec:dicsussions:xr}.

These results align with and extend previous findings from VRTest~\citep{wangVRTestExtensibleFramework2022}. Its greedy approach, VRGreed, achieved $\sim$80\% coverage for interactable objects across five VR scenes (with 100\% in two scenes), while its pure random VRMonkey achieved only $\sim$10\% (0\% in three scenes).
Our work amplifies this effectiveness gap, as our subject VR apps involve more complex spatial interactions that require precise navigation and manipulation, making random approaches particularly ineffective at locating and engaging with target interactables.

\begin{rqanswer}
\textbf{Answer to RQ1:} Across 373 \emph{fire}, \emph{manipulate}, and \emph{socket} interactions, \TOOL demonstrated great effectiveness with 93\% coverage, vastly outperforming the random baseline which achieved only 8\% coverage overall. 
\end{rqanswer}

\subsection{RQ2: Efficiency} \label{sec:evaluation:effeciency}

\begin{figure}[t]
    \centering
    \includegraphics[width=1\linewidth]{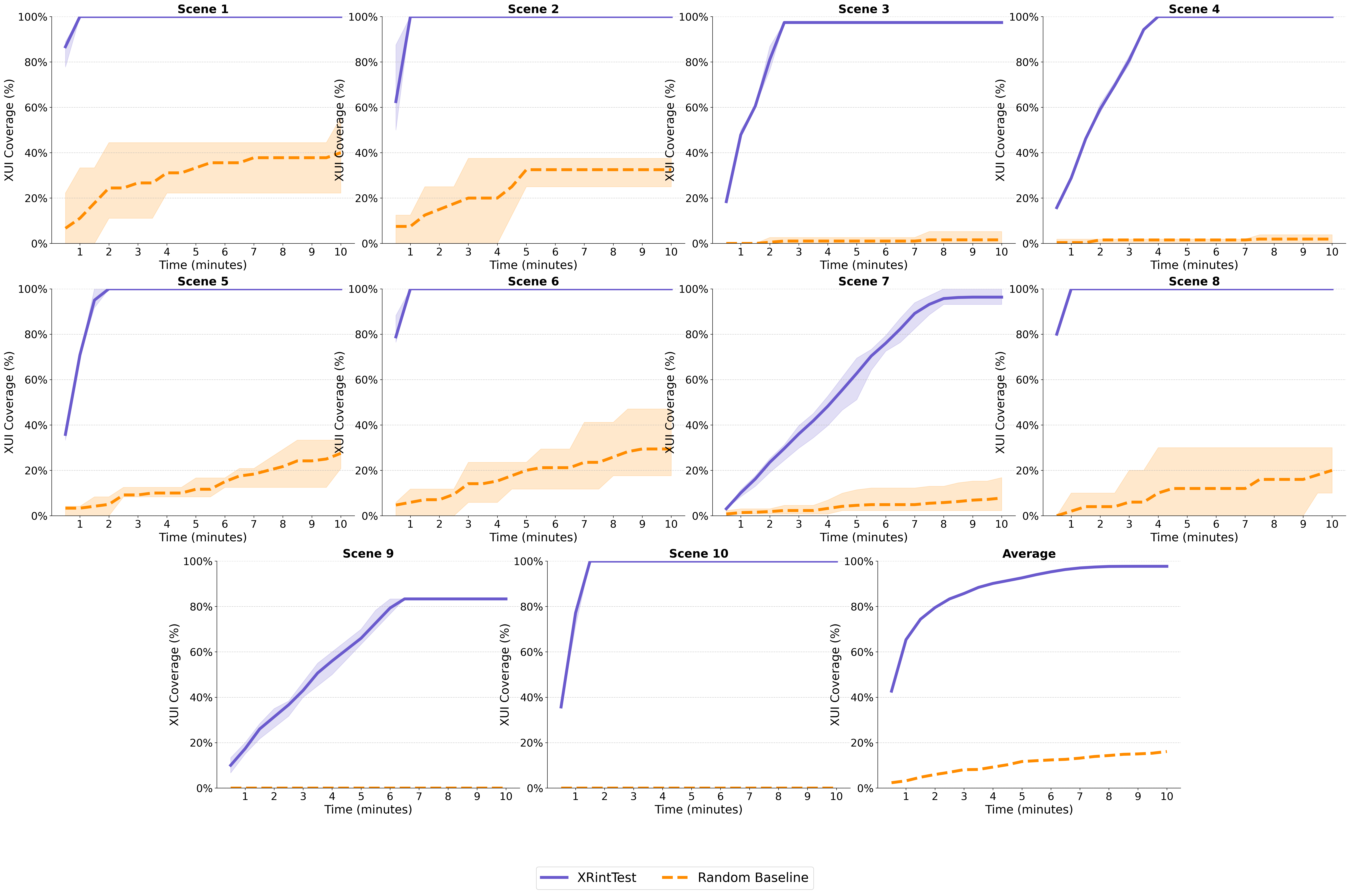}
    \caption{Efficiency of \TOOL in Achieving VR User Interaction Coverage}
    \label{fig:results:efficiency}
\end{figure}

Figure~\ref{fig:results:efficiency} compares the efficiency of \TOOL and the random baseline in achieving $IFC$ across seven subject scenes, averaged from five executions, and the last chart shows the averaged results for all scenes. 
The results show \TOOL's high efficiency across all subject scenes. Notably, in scenes 1, 2, 5, 6, 8, and 10, it rapidly achieved nearly 100\% $IFC$ within the first two minutes of the test sessions. 
For the rest scenes, we observe a more gradual coverage gain due to the significantly larger number of interactions they contain. However, \TOOL still completed the test sessions and achieved high coverage within the allocated time budget.

In contrast, the random baseline exhibited slower interaction coverage efficiency. While showing continuous improvement over time in several scenes, it struggles to achieve over 40\% coverage in the allocated time. 
The last averaged chart illustrates this significant difference in efficiency, with \TOOL reaching approximately 95\% coverage at the midpoint of the allocated time while the random baseline achieves less than 15\% coverage in the same timeframe.

The steeper coverage curves of \TOOL across all scenes highlight its ability for rapid testing of the interactions in VR scenes. The efficiency advantage is particularly important for testing large, complex scenes (e.g., scene 7 in our dataset), with high spatial complexity and interaction density. 

\begin{rqanswer}
\textbf{Answer to RQ2:} \TOOL achieves high coverage within 2 minutes in most scenes, reaching near full coverage within the allocated time for all scenes, consistently outperforming the random baseline across subjects. 
\end{rqanswer}

\subsection{RQ3: Bug Detection} \label{sec:evaluation:bug}
 
%
Both \TOOL and the random baseline successfully detected the same exception in scene 3, demonstrating comparable ability in identifying runtime errors. 
We traced the exception to an implementation issue and reported it to the project's maintainers.
Furthermore, \TOOL's test result checker (Section~\ref{sec:approach:oracle}) identified two unresponsive interaction issues among the failed \emph{manipulate} interactions for scene 7. 
Notably, one of the unresponsive interactions \emph{had not been identified during our manual exploration of subject scenes}. This finding highlights \TOOL's bug detection capabilities, which in this instance surpassed human testing efforts.

Our root cause analysis revealed a key distinction between VR interaction issues. While the runtime exception in scene 3 stemmed from an implementation error in the source code, the two unresponsive interaction issues resulted from misconfigured object properties. 
Specifically, these issues were caused by misconfiguration of the objects' \emph{collider} properties (Section~\ref{sec:background:unity}). 
These interactables had their \texttt{Collider.isTrigger} parameter enabled, which prevents physical collision and instead enables specific event triggering (e.g., function calls, typically for UI elements).
Consequently, these objects could not physically engage with other objects (e.g., controllers), rendering them impossible to interact with despite having proper interaction definitions.
To ensure validity, we are communicating with the respective VR project developers to confirm the identified issues\footnote{GitHub issue: \url{https://github.com/mikeNspired/XRI-Starter-Kit/issues/36}}.

These findings underscore the importance of testing approaches that can identify configuration-related defects in addition to traditional code-level or runtime exception bugs. This is especially important for VR apps, where physics-based interactions and realism are fundamental to user experience, but effective exploratory testing is challenging for human testers.

\begin{rqanswer}
\textbf{Answer to RQ3:} Both \TOOL and the random baseline detected the same runtime exception, but only \TOOL found two additional \emph{unresponsive interaction} bugs attributable to misconfigured object properties, one of which had been missed during manual exploration.
\end{rqanswer}

\subsection{Threats to Validity}


A potential threat to external validity lies in the selection of VR applications used in our evaluation.
While our approach focuses primarily on Unity-based VR implementations, we argue that the underlying principles and findings are broadly generalisable across XR (i.e., VR, AR, and MR) ecosystems.
Unity currently dominates VR development with over 60\% market share~\citep{wangVRGuideEfficientTesting2023} and offers integrations with virtually all major XR platforms~\citep{roberts2023ARVRTechnology}. 
Furthermore, Unity remains the most widely adopted platform for evaluating XR software testing research~\citep{gu2025softwaretestingextendedreality}.
These factors together make Unity-based projects a representative and practical foundation for developing and validating our approaches.

While our benchmark \DATASET contains tutorial and sample VR apps, potentially raising representativeness concerns, we mitigate this by manually verifying that the subjects encompass diverse VR interaction patterns and represent core mechanisms used in modern VR apps.
Specifically, scenes 3, 4, 7, and 10 each feature over 50 distinct interactions, including numerous custom patterns and complex testing scenarios. 
Furthermore, scenes 7 and 9 are both full VR puzzle-solving games that require exploration and specific object interactions to complete the game level, demonstrating intricate scene functionality.
Notably, VR is an emerging field, and the availability of open-source subjects is currently limited~\citep{gu2025XRGlasses}. We expect future replication studies as more open-source VR apps become available.



A potential threat to internal validity is that our approach and evaluation are based on the ``interactor-interactable'' user interaction pattern defined by Unity XRI in the context of VR application development.
However, the conceptual design of our approach transcends both specific XR technologies (e.g., VR) and particular development platforms (e.g., Unity). 
The generalisation of our method can be discussed from two complementary perspectives: \emph{conceptual} and \emph{technical}.

At the \emph{conceptual level}, our approach is founded on the ``interactor-interactable'' user interaction pattern, which serves as the prevailing model for managing user input across modern XR apps. 
Within this paradigm, users employ input devices to perform actions on objects within a spatial environment, which is foundational not only to VR but also to AR and MR systems. 
Consequently, our approach should remain applicable to any XR framework adhering to this interaction model, regardless of platform or hardware variation.

At the \emph{technical aspect}, the most direct parallel platform beyond Unity is with Unreal Engine, which is the other leading engine for XR development and employs a conceptually identical paradigm. 
Unreal Engine\footnote{\url{https://www.unrealengine.com/xr}}'s \texttt{Motion Controller} component\footnote{\url{https://dev.epicgames.com/documentation/unreal-engine/motion-controller-component-setup-in-unreal-engine}} manages input from XR input devices much like Unity's ``Interactor''.
Furthermore, objects are made interactive by adding components like the \texttt{Grab Component}\footnote{\url{https://dev.epicgames.com/documentation/unreal-engine/vr-template-in-unreal-engine\#grab}}, which are analogous to Unity XRI's \texttt{XR Grab Interactable}.
Therefore, porting our approach to Unreal Engine would primarily involve a systematic mapping between engine-specific APIs and constructs, rather than a redesign of the underlying logic or abstractions. The foundational principles of our approach would remain unchanged under such a translation.

Additionally, the same principles could be directly extended beyond VR to other XR modalities, including AR and MR. 
In fact, Unity's XR Interaction Toolkit, which is a primary subject of our work, is directly applicable for building AR/MR applications\footnote{\url{https://docs.unity3d.com/Packages/com.unity.template.mixed-reality@2.1/manual/index.html}}.
Although with different XR technologies, the interaction patterns (e.g., grabbing, triggering, and controller movement) are analogous.
Our decision to evaluate only VR applications was therefore motivated by pragmatic considerations, particularly the availability of publicly accessible open-source projects suitable for automated testing. At the time of evaluation, our search did not identify AR/MR projects with sufficiently rich interaction structures to support meaningful experimentation, and even Unity's official MR template contained only a single interactable object. This limitation thus reflects the scarcity of comprehensive research artefacts rather than a conceptual constraint of our method.


The current work represents an initial step in this direction. Future studies will aim to rigorously evaluate our solutions' generalisation capabilities across broader XR modalities and adapt them to evolving patterns of spatial interaction.

%% file: sections/8-discussion.tex
\section{Discussion} \label{sec:dicsussions}


\subsection{Interaction Design Smells} \label{sec:dicsussions:smell}

Design smells indicate potential problems in software. For instance, code smells and test smells signal issues in source and test code~\citep{Liu2016CodeSmell, Spadini2018TestSmells}. Similarly, UI design smells violate guidelines in complex systems, such as Android apps~\citep{Yang2021UIS-Hunter}, often involving elements that can be dismissed unexpectedly or hinder access to key information.

In our findings, we consider setting socket interactors' Interaction Layer Masks to ``Everything'' as a potential interaction design smell that violates intended functionalities. 
We are engaging with developers of affected VR projects to confirm this as a bad practice\footnote{GitHub issue: \url{https://github.com/Unity-Technologies/XR-Interaction-Toolkit-Examples/issues/140}}.

Furthermore, we found that the IFG can be used to effectively detect this type of interaction design smells. 
The IFG identify interaction brokerages \ref{sec:approach:xui_graph:brokerage} by matching interactables and socket interactions sharing the same Interaction Layer Masks. 
Normal interactables (i.e., those not paired with sockets) typically use the ``Everything'' layer.
When socket interactors also default to ``Everything'', the IFG generates an explosion of unintended edges, such as linking all nodes to a ``Door'' interactable in Figure~\ref{fig:xui_graph}.
This misconfiguration could render the graph excessively dense and uninformative.

\subsection{Oracle Automation} \label{sec:dicsussions:oracle}

Test oracle automation represents a major challenge in VR automated testing~\citep{gu2025softwaretestingextendedreality}. 
The current implementation of \TOOL employs a contract-based oracle automation approach that validates object properties as postconditions~\citep{molinaTestOracleAutomation2025}. Its generalised mechanism has been proven effective without requiring manual intervention or scene-specific customisation.
While effective, this approach can only validate interaction activation, missing semantic interaction failures where interactions technically succeed but produce incorrect behaviours. For instance, a movable object may respond to grab actions but cannot actually be moved \emph{as intended}, representing an opportunity for improvement.

The complexity and variability of VR interactions make it particularly challenging to determine expected outcomes programmatically. While metamorphic testing has shown promising results in assessing VR runtime requirements~\citep{deandradeExploitingDeepReinforcement2023}, our evaluation revealed the difficulty of defining meaningful metamorphic relationships (MRs) for diverse interaction patterns and their resulting behaviours.
The challenge lies not only in identifying appropriate MRs but also in efficiently validating them across the wide spectrum of possible VR app behaviours. 

\subsection{Environment-Aware Testing} \label{sec:dicsussions:xr}

\TOOL currently focuses on intended user interactions, and our evaluation demonstrates effective detection of direct interaction failures (Section~\ref{sec:evaluation:bug}).
While testing Scene 7, we observed instances where interactions became permanently inaccessible after being destroyed through physics-based interactions inadvertently caused by test actions. 
Similarly, for Scene 10, the reason behind the relatively lower coverage for \emph{socket} interactions compared to other types of interactions is that the paired interactables for the socket interactors became permanently inaccessible, making those socket interactions impossible to cover.

The challenge lies in \TOOL's current focus on isolated interaction validation without considering the cascading effects of actions within the VR environment. While our oracle successfully detects when individual interactions fail, it cannot predict or prevent scenarios where successful interactions lead to environmental changes that compromise subsequent testing. 
For instance, a physics simulation that correctly responds to user input may inadvertently destroy other interactable objects.
This limitation suggests the need for more comprehensive environment modelling that also accounts for the complex interplay between objects, physical systems, and scene-wide changes that characterise realistic.



%% file: sections/9-conclusion.tex
\section{Related Work} \label{sec:related_work}

\subsection{Video Game Testing}

\citet{politowski2022AutomatedVideoGame} conducted a literature review on automated video game testing and an online survey with game developers. The review shows increasing research interest in automated game testing, while the survey indicates that many developers remain sceptical about using automated agents in practice. 
\citet{zheng2019WujiAutomaticOnline} developed \textsc{Wuji}, a framework that automatically tests games using evolutionary algorithms and reinforcement learning to explore game spaces, traverse branches, and progress through stages.
\citet{Chen2021GLIB} proposed \textsc{GLIB}, a code-based data augmentation technique that automatically detects game GUI glitches. 
\citet{Prasetya2022iv4xrReport} presented \textsc{iv4XR}, a multi-agent programming framework that supports game testing by enabling test agents to interact with the game under test.
Video games are typically goal-driven, requiring players to complete specific tasks or levels. In contrast, XR applications are often not game-like and do not share the same testing assumptions or strategies as video games.

\subsection{Mobile Application Testing}

\citet{suGuidedStochasticModelbased2017} proposed \textsc{Stoat}, a stochastic model-based testing tool for Android that combines static and dynamic analysis to systematically generate event sequences.
\citet{Xiong2024Kea} introduced \textsc{Kea}, a property-based testing framework designed to detect functional bugs by validating app behaviours against user-defined invariants.
\citet{Chen2024HarmonyOS} developed a model-based GUI testing approach that constructs a page transition graph through static analysis to guide systematic exploration.
\citet{Yoon2024IntentGUILLM} presented \textsc{DroidAgent}, an LLM-powered testing tool that leverages language models to define testing goals and autonomously interact with mobile interfaces to achieve them.
While these approaches demonstrate significant progress in automated GUI testing for mobile platforms, they remain limited to 2D graphical interfaces and lack the spatial, interactive characteristics of modern XR environments.

\subsection{XR Software Testing}

\citet{Li2020WebXR} conducted the first empirical study of WebXR bugs, analysing 368 real bugs from 33 GitHub-hosted projects through seven rounds of manual classification to build a taxonomy based on symptoms and root causes, and comparing them to bugs in conventional JavaScript and web applications.
\cite{li2024LessCybersicknessPlease} introduced \textsc{StereoID}, a black-box testing framework that detects stereoscopic visual inconsistencies (known as cybersickness) solely from rendered GUI states, addressing rendering flaws that cause user discomfort.
\citet{Qi2025LLM4VR} explored the use of large language models for field-of-view (FOV) analysis in VR exploration testing through a dedicated case study.
\citet{gu2025softwaretestingextendedreality} presented the first systematic mapping study on XR software testing, analysing 34 studies on techniques and empirical approaches, classifying them by research landscape, test facets, and evaluation methodologies.

\section{Conclusion and Future Work} \label{sec:conclusion}

This paper addresses key challenges in VR app testing by systematically modelling interactions within VR scenes and automatically generating 3D user interactions.
We first empirically analysed seven diverse VR scenes in \DATASET, identifying 367 user interactions spanning four distinct categories (\emph{fire}, \emph{manipulate}, \emph{socket} and \emph{custom}), which characterise prevalent interaction patterns.
We then introduced the \emph{Interaction Flow Graph} (IFG), a model representing 3D user interactions within VR scenes, which underpins \TOOL, an automated testing tool that systematically explores VR scenes and activates identified interactions. 
Evaluation on \DATASET shows that \TOOL achieves 93\% interaction flow coverage, outperforming a random baseline by a factor of 12 in effectiveness and six in efficiency, and can detect VR interaction issues. 

Although our approach effectively handles fundamental VR interactions, various custom patterns remain unexplored.
We propose a two-phase future work, starting with a systematic study to characterise common custom interaction patterns and challenges they pose for automated testing (e.g., defining robust test oracles).
We aim to integrate deep reinforcement and imitation learning (RL/IL) techniques to adaptively address these patterns.
RL/IL has proven effective for complex object manipulation in fields like robotics~\citep{Xie2020Bimanual}, suggesting potential directions.
Additionally, RL/IL has already demonstrated success in testing domains of Android GUI~\citep{Romdhana2022DeepBlackAndroid} and video games~\citep{Zheng2019Wuji}.
Building on \TOOL's foundational framework, we provide a clear pathway for extending automated 3D VR testing with advanced learning-based methods to meet the challenges of future immersive environments.